\documentclass[aps,pra,twocolumn,superscriptaddress,floatfix,longbibliography]{revtex4}
\usepackage{amsfonts}
\usepackage{amssymb}
\usepackage{graphicx}
\usepackage{dcolumn}
\usepackage{bm}
\usepackage{amsmath}
\usepackage[colorlinks,linkcolor=magenta,citecolor=blue,urlcolor=blue]{hyperref}
\usepackage{changes}
\begin{document}

\title{Mobility edges and critical regions in periodically kicked incommensurate optical Raman lattice}
\author{Yucheng Wang}
\thanks{Corresponding author: wangyc3@sustech.edu.cn}
\affiliation{Shenzhen Institute for Quantum Science and Engineering,
Southern University of Science and Technology, Shenzhen 518055, China}
\affiliation{International Quantum Academy, Shenzhen 518048, China}
\affiliation{Guangdong Provincial Key Laboratory of Quantum Science and Engineering, Southern University of Science and Technology, Shenzhen 518055, China}
%\Letter Yucheng Wang \\
%\email{wangyc3@sustech.edu.cn}}

\begin{abstract}
Conventionally the mobility edge (ME) separating extended states from localized ones is a central
concept in understanding Anderson localization transition. The critical state, being delocalized and non-ergodic, is a third type of fundamental state that is different from both the extended and localized states. Here we study the localization phenomena in a one dimensional periodically kicked quasiperiodic optical Raman lattice by using fractal dimensions. We show a rich phase diagram including the pure extended, critical and localized phases in the high frequency regime, the MEs separating the critical regions from the extended (localized) regions, and the coexisting phase of extended, critical and localized regions with increasing the kicked period. We also find the fragility of phase boundaries, which are more susceptible to the dynamical kick, and the phenomenon of the reentrant localization transition.
Finally, we demonstrate how the studied model can be realized based on current cold atom experiments and how to detect the rich physics by the expansion dynamics. Our results provide insight into studying and detecting the novel critical phases, MEs, coexisting quantum phases, and some other physics phenomena in the periodically kicked systems.
\end{abstract}

%\pacs{74.20.-z, 74.78.-w, 05.30.Rt, 71.10.Pm}
%74.20.-z:theories and models of superconducting state; 74.78.-w:superconduting films and low-dimensional structures;
%05.30.Rt: quantum phase transitions; 71.10.Pm: fermions in reduced dimensions .
\maketitle
%%%%%%%%%%%%%%%%%%%%%%%%
\section{Introduction}
%%%%%%%%%%%%%%%%%%%%%%%%%
Anderson localization (AL)~\cite{Anderson1958,RMP1,RMP2}, namely that eigenfunctions are exponentially localized in space because of the quantum interference in disordered systems, is a fundamental quantum phenomenon in nature. The transition between metal (extended) phase and insulator (localized) phase can occur for sufficiently strong disorder in three dimensional systems, and near the transition point, there exist mobility edges (MEs) which mark the critical energy separating the extended and localized states~\cite{Lagendijk2009,Evers2008}.
MEs lie at the heart of studying various fundamental localization phenomena such as the disorder induced metal-insulator transition. Since the effect of suppressing diffusion via quantum interference is particularly pronounced in one and two dimensions, in which the eigenstates are always localized for arbitrarily small disorder strengths~\cite{Anderson1979}, and thus, no MEs exist. Besides the random disorder, quasiperiodic potentials can also induce the extended-AL transition, and bring about different physics, e.g., the existence of Anderson transition and MEs even in one dimensional (1D) systems~\cite{AA,Xie1988,Biddle0,Biddle,Pu2013,Pin2014,Ganeshan2015,Santos2019,YaoH2019,YuWang2020,JBiddle,XiaoLi,YuchengME,Basu2021,TLiu2022,Ribeiro} and multifractal critical states~\cite{Hatsugi1990,Takada2004,Liu2015,YuchengC1,Cai2013,WangYC2016,WangJ2016,YuchengC2,BoYan2011}.
Critical phase is a third type of phases, and is fundamentally different from the localized and extended phases
in the spectral statistics~\cite{Geisel,Jitomirskaya}, wave functions' distributions~\cite{Halsey,Mirlin}, and dynamical behaviors~\cite{Hiramoto,Ketzmerick}.

Quasiperiodic systems have been realized in ultracold atomic gases by superimposing two 1D optical lattices with incommensurate wavelengths, and the extended-localized transition and MEs have been observed~\cite{Roati2008,BlochME,Gadway2018,BlochME2,Gadway2020}.
%In this system, The existence of the many body localization phase in the AA model in the presence of weak interactions has also been well established in both theory~\cite{DAHuse2013,YuWang,Das1902} and experiment~\cite{Bloch1,Bloch3}.
However, the critical phase has not been strictly realized in experiment until now.
In recent works, we have proposed to realize the critical phase in the optical Raman lattice~\cite{YuchengC2}, which possesses the spin-orbit coupling term and an incommensurate Zeeman potential. Further, we have predicted a coexisting phase consisting of three different energy-dependent regions, i.e., the extended, localized and critical regions~\cite{YuchengC3}, which shows the abundant transport features. Recently, T. Shimasaki et. al. reported the experimental observation of the critical states and anomalous localization in a kicked quasiperiodic Aubry-Andr\'{e} (AA) lattice~\cite{Shimasaki2022}. In the kicked AA model, there is not the critical phase but the phase with coexisting critical and localized (or extended) regions~\cite{YuZhang}. An important question is whether the critical phase consisting of solely critical eigenstates and the most nontrivial coexisting phase composed of three different regions can be realized in kicked systems.

Motivated by the recent experimental realizations of the optical Raman lattices~\cite{LiuXJ2013,LiuXJ2014,Lepori2016,WangBZ2018,Liu2016,Song2018,Song2019,JWPan,RamanReview2018} and kicked systems with AL in ultracold atomic gas~\cite{Shimasaki2022,Weld2022}, we propose a scheme to realize the critical phase and the coexisting phase based on the 1D optical Raman lattice with periodically kicked quasiperiodic Zeeman potential. This system displays extremely rich localization phenomena as the change of the driven period.
%%%%%%%%%%%%%%%%%%%%%%%%%%%%%%%%%%%%%%%%%%%%%%%
\begin{figure*}[t]
\hspace*{-0.6cm}
\includegraphics[width=0.95\textwidth]{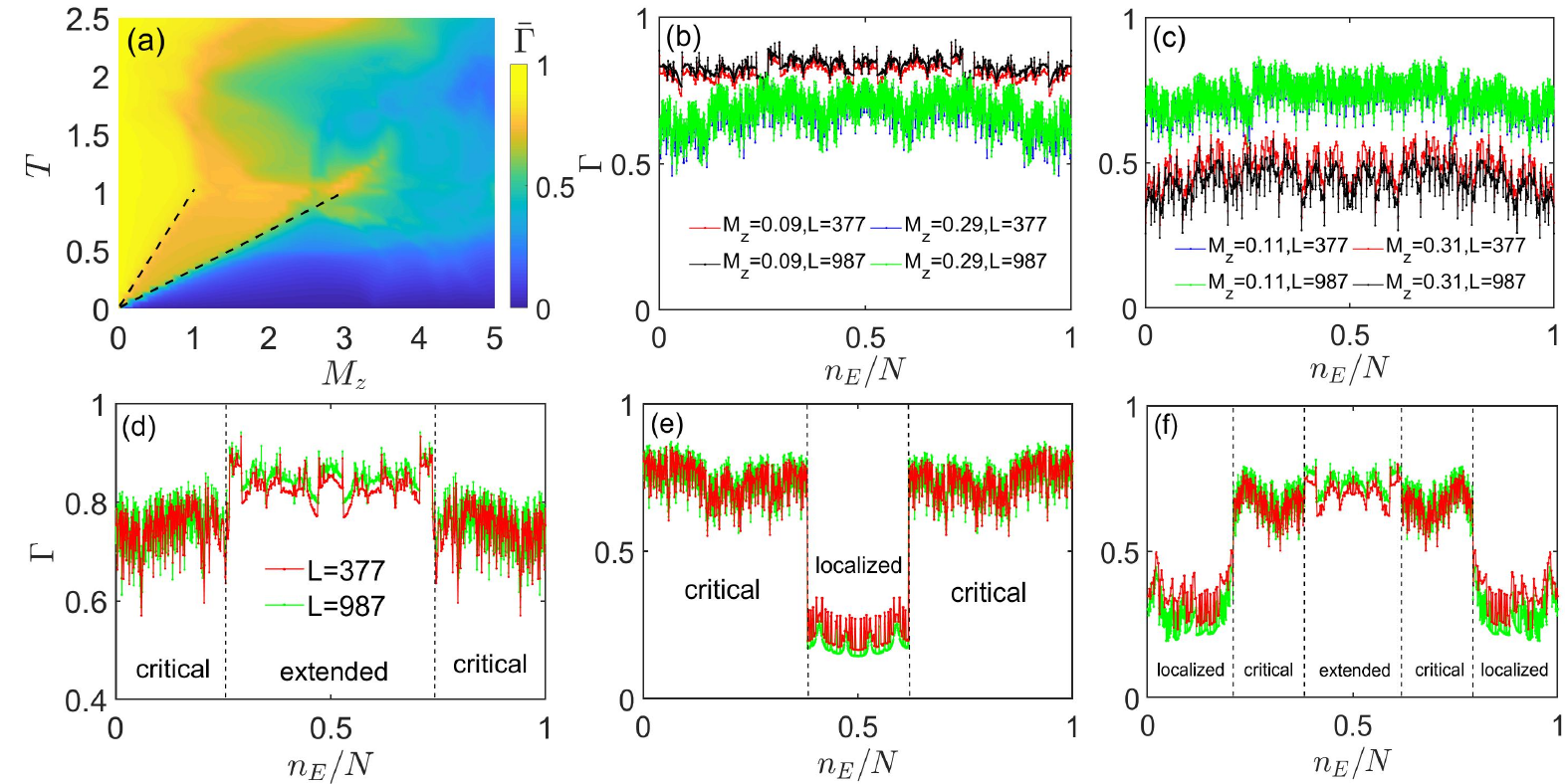}
\caption{\label{01}
 (a) The mean fractal dimension $\bar{\Gamma}$ with $L=F_{14}=610$ as a function of the kicked quasiperiodic potential strength $M_z$ and the period $T$. The dashed lines correspond to $M^c_z/T=2|J_0-J_{so}|$ and $M^c_z/T=2(J_0+J_{so})$, respectively. $\Gamma$ for the wave-function of each mode at (b) $M_z=0.09, 0.29$, (c) $0.11, 0.31$ with different sizes and fixed $T=0.1$. The index of energy mode $n_E$ runs from $1$ to $N=2L$. $\Gamma$ as a function of $n_E/N$ for the sizes $L=F_{13}=377$ and $L=F_{15}=987$ with (d) $M_z=0.8$, $T=0.8$, (e) $M_z=3$, $T=0.8$, and (f) $M_z=2.3$, $T=1.7$. Here we set $J_0=1$ and $J_{so}=0.5$.}
\end{figure*}
%%%%%%%%%%%%%%%%%%%%%%%%%%%%%%%%%%%%%%%%%%%%%%%%

%%%%%%%%%%%%%%%%%%%%%%%%%%%%%%%%%%%%%%%%
\section{Model and phase diagram}
%\label{non-interacting}
%%%%%%%%%%%%%%%%%%%%%%%%%%%%%%%%%%%%%%%
We propose the periodically kicked quasiperiodic optical Raman lattice model described by
\begin{equation}\label{Hsum}
H=H_0+H_{\rm SOC}+H_{K},
\end{equation}
with
\begin{subequations}
\begin{eqnarray}
H_0= -J_0\sum_{\langle i\rangle}(c^{\dagger}_{i,\uparrow}c_{i+1,\uparrow}-c^{\dagger}_{i,\downarrow}c_{i+1,\downarrow})+H.c., \quad\\
%\label{Lambda7a}
H_{\rm SOC}=J_{so} \sum_i(c^{\dagger}_{i,\uparrow}c_{i+1,\downarrow}-c^{\dagger}_{i,\uparrow}c_{i-1,\downarrow})+H.c.,  \quad\\
H_{K}=\sum_n\delta(t-nT)\sum_{i}\mu_i(n_{i,\uparrow}-n_{i,\downarrow}), \quad \qquad
%\notag
\end{eqnarray}
\end{subequations}
where $c_{i,\sigma}$, $c^{\dagger}_{i,\sigma}$ and $n_{i,\sigma}=c^{\dagger}_{i,\sigma}c_{i,\sigma}$ are the annihilation, creation and particle number operators at lattice site $i$, respectively, and $\sigma=\uparrow,\downarrow$ denotes the spin.
The term $H_0$ ($H_{\rm SOC}$) presents the nearest neighbor spin-conserved (spin-flip) hopping with strength $J_0$ ($J_{so}$), and for convenience, we set $J_0=1$ as the energy unit. $H_K$ denotes the kicking part with
\begin{equation}
 \mu_i=M_z\cos(2\pi\alpha i+\phi),
\label{mu}
\end{equation}
where $\alpha$ and $\phi$ are the irrational number and phase shift, respectively. Without loss of generality, we set
$J_{so}=0.5$, $\phi=0$~\cite{explain1,explain2} unless otherwise stated, and $\alpha=(\sqrt{5}-1)/2$, which is approached by $\alpha=\lim_{m\rightarrow\infty}\frac{F_{m-1}}{F_m}$. Here $F_m$ is the Fibonacci number defined by $F_{m+1}=F_{m-1}+F_{m}$ with the starting values $F_0=F_1=1$~\cite{Kohmoto1983}. For a finite system with size $L=F_m$, we take $\alpha=\frac{F_{m-1}}{F_m}$ when using periodic boundary conditions. When the Zeeman potential is constantly turned on, i.e., $H_K=\sum_{i}\mu_i(n_{i,\uparrow}-n_{i,\downarrow})$, there are three distinct phases: extended, critical and localized phases~\cite{YuchengC2}. The phase boundary between the extended and critical phases satisfies $M^c_z=2|J_0-J_{so}|$ and the phase boundary between the critical and localized phases satisfies $M^c_z=2(J_0+J_{so})$.

The dynamical evolution of this kicked system is described by the Floquet unitary propagator over one period, i.e.,
\begin{equation}
 U(T)=e^{-i(H_0+H_{\rm SOC})T}e^{-i\sum^L_{j=1}\mu_j(n_{j,\uparrow}-n_{j,\downarrow})}.
\label{Floquet}
\end{equation}
Here we have set $\hbar=1$. In the basis of $|j,\sigma\rangle$, $\langle i,\sigma|U|j,\sigma'\rangle$ is a $2L\times 2L$ matrix.
For a initial state $|\psi(0)\rangle$, the evolution state after $N_K$ kicked periods is given by $|\psi(N_KT)\rangle=[U(T)]^{N_K}|\psi(0)\rangle$. Thus, the distribution of the eigenstate $|\psi_{\beta}\rangle$ of the
propagator $U(T)$ with the Floquet energy $E_{\beta}$, i.e., $U(T)|\psi_{\beta}\rangle=e^{-iE_{\beta}T}|\psi_{\beta}\rangle$, can reflect the dynamical property of this kicked system. To describe the distribution, we introduce the fractal dimension, which for an arbitrary eigenstate $|\psi_{\beta}\rangle=\sum_{j=1}^L[u_{\beta,j}c^{\dagger}_{j,\uparrow}+v_{\beta,j}c^{\dagger}_{j,\downarrow}]|0\rangle$ is defined as
\begin{equation}
 \Gamma=-\lim_{L\rightarrow\infty}\frac{\ln(IPR)}{\ln L},
\label{mu}
\end{equation}
where $IPR=\sum_{j=1}^L(u^4_{\beta,j}+v^4_{\beta,j})$ is the inverse participation ratio (IPR)~\cite{RMP1}. It is known that $\Gamma\rightarrow 0 (1)$ for the localized (extended) states, while $0<\Gamma<1$ for the critical state. To sketch out the phase diagram, we define the mean fractal dimension over all eigenstates: $\bar{\Gamma}=(2L)^{-1}\sum^{2L}_{\beta=1}\Gamma(\beta)$. Fig.~\ref{01} (a) shows $\bar{\Gamma}$ as a function of $M_z$ and $T$ with fixed $J_{so}=0.5$. In the high-frequency regime $T\ll 1$, it is shown that the phase boundaries between the critical and extended or localized phases of this system can be well described by the dashed lines, which correspond to
\begin{eqnarray}\label{boundary}
%\begin{equation}
M_{z}^{c}/T= \qquad \qquad \qquad \qquad \qquad \qquad \qquad \qquad \qquad \qquad \nonumber\\
\begin{cases}
2|J_0-J_{so}|, \textrm{between extended and critical phases}, \\
2(J_0+J_{so}), \textrm{between critical and localized phases}.
\end{cases}
%\end{equation}
\end{eqnarray}
To see it clearly, in Figs.~\ref{01} (b) and (c), we fix $T=0.1$ and show $\Gamma$ of different eigenstates as a function of $n_E/N$ for different sizes, where $N=2L$ is the number of the total eigenstates. One can observe that $\Gamma$ tends to $1$ for all states at $M_z=0.09$ (satisfying $M_z/T< 2|J_0-J_{so}|$) with increasing the system size, meaning that they are extended, while $\Gamma$ tends to $0$ for all states at $M_z=0.31$ (satisfying $M_z/T> 2(J_0+J_{so})$) when increasing the size, implying that all states are extended. In contrast, when $M_z=0.11$ and $0.29$ (satisfying $2|J_0-J_{so}|< M_z/T< 2(J_0+J_{so})$), $\Gamma$ is clearly different from $0$ and $1$, and almost independent of the system size, showing that all states are critical. To understand this result, we derive the effective Hamiltonian $H_{eff}$ in the high-frequency regime, namely
\begin{equation}
 U(T)=exp(-iH_{eff}T).
\label{Ut}
\end{equation}
By using the Baker-Campbell-Hausdorff formula~\cite{BCH} and combining Eq.~(\ref{Floquet}) and Eq.~(\ref{Ut}), one can obtain,
\begin{eqnarray}
H_{eff}=H_0+H_{\rm SOC}+\frac{M_z}{T}V-i\frac{M_z}{2}[H_0+H_{\rm SOC}, V] \nonumber\\
+\frac{TM_z}{12}[H_0+H_{\rm SOC}, [H_0+H_{\rm SOC}, V]]+\cdots,
\label{Lambda6}
\end{eqnarray}
where $V=\sum^{L}_{j=1}\cos(2\pi\alpha j)(n_{j,\uparrow}-n_{j,\downarrow})$. When $1/T\gg 1$ and $M_z\ll 1$, the effective Hamiltonian is simplified as $H_{eff}=H_0+H_{\rm SOC}+\frac{M_z}{T}V$, which is equivalent to that obtained by transforming the $H_K$ in Eq.~(\ref{Hsum}) into
$H_K=\sum_{i}\mu_i/T(n_{i,\uparrow}-n_{i,\downarrow})$ and leaving $H_0$ and $H_{\rm SOC}$ unchanged. Compared with the
non-kicked case~\cite{YuchengC2}, this effective model includes three distinct phases, and the phase boundaries satisfy Eq.~(\ref{boundary}).

%%%%%%%%%%%%%%%%%%%%%%%%%%%%%%%%%%%%%%%%%%%%%%
\begin{figure}[t]
\hspace*{-0.2cm}
\includegraphics[width=0.49\textwidth]{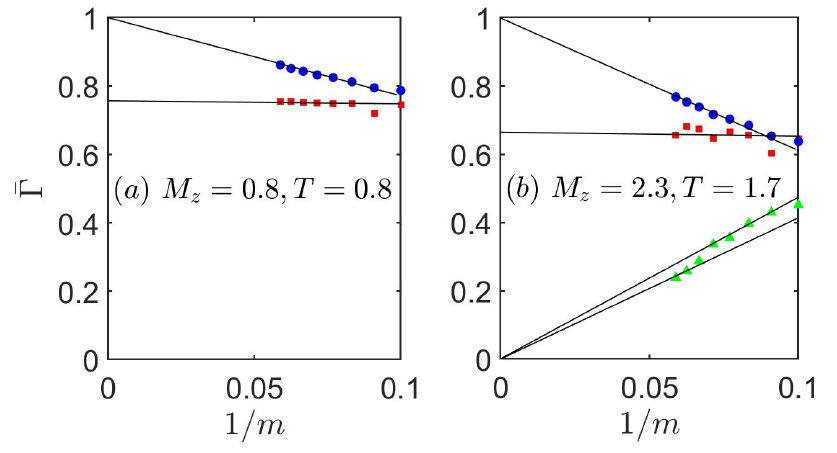}
\caption{\label{02S}
$\bar{\Gamma}$ as a function of $1/m$ for different regions with (a) $M_z=0.8$, $T=0.8$ (corresponding to Fig.~\ref{01} (d)), and (b) $M_z=2.3$, $T=1.7$ (corresponding to Fig.~\ref{01} (f)). Here $m$ are the Fibonacci indices. }
\end{figure}
%%%%%%%%%%%%%%%%%%%%%%%%%%%%%%%%%%%%%%%%%%%%%%%%

With increasing $T$, the high-order terms in Eq.~(\ref{Lambda6}) can't be neglected, and thus, the effective Hamiltonian includes the non-neighbor hopping term, which will induce the occurrence of MEs~\cite{Biddle0,Biddle,Santos2019}. Figs.~\ref{01} (d) and (e) show $\Gamma$ for different sizes and $M_z$ with fixed $T=0.8$. We see that $\Gamma$ tends to $1$ ($0$) for the states in center of energy spectra of the system with $M_z=0.8$ ($M_z=3$) when increasing the system size, suggesting that they are extended (localized). In contrast, in the tails of the energy spectra in both Figs.~\ref{01} (d) and (e), the fractal dimension $\Gamma$ is clearly different from $1$ and $0$ , and is almost
independent of system sizes, implying that all states are critical. Thus, there exist energy-dependent extended and critical regions when $M_z=0.8$, and energy-dependent localized and critical regions when $M_z=3$, meaning that there are the MEs separating the extended and localized states from the critical states, which are different from the conventional MEs separating the extended states from the localized ones. With the further increasing of $T$, there will be the quantum phase with three coexisting fundamentally different regions, i.e., the localized, extended, and critical regions, as shown in Fig.~\ref{01} (f).

Now we consider the finite size effect of the fractal dimension $\Gamma$. When changing the system size, the number and magnitudes of the eigenvalues will change accordingly. Thus, it is difficult to carry out the finite size scaling analysis for a fixed eigenstate. We take a coarse graining on the spectrum and investigate the average $\Gamma$ over the eigenstates in a single region, and accordingly we define
\begin{equation}
 \bar{\Gamma}=\frac{1}{N_r}\sum_{same\ region}\Gamma,
\label{meanG}
\end{equation}
where $N_r$ is the number of eigenstates in the region and can be obtained by comparing the fractal dimension with different sizes, as shown in Figs.~\ref{01} (b-f). Since all eigenstates in the same region have the same properties, the average fractal dimension in an
arbitrary small sub-region of a region can also be similarly defined, and will display the same scaling behavior with the region. Figs.~\ref{02S} (a) and (b) show $\bar{\Gamma}$, which are obtained by computing the average $\Gamma$ of all states in the same region of the systems corresponding to Figs.~\ref{01} (d) and (f), respectively. $\bar{\Gamma}$ extrapolates to $1$ and $0.75$ in the extended and critical regions of Fig.~\ref{01} (d), which confirms that the corresponding states in these regions are extended and critical, respectively. $\bar{\Gamma}$ respectively extrapolates to $0$, $1$ and the value far from $0$ and $1$ in the three different regions of Fig.~\ref{01} (f), which confirms the corresponding system with three coexisting energy-dependent regions, i.e., the extended, localized, and critical regions.

%%%%%%%%%%%%%%%%%%%%%%%%%%%%%%%%%%%%%%%%
\begin{figure*}[t]
%\hspace*{-0.3cm}
\includegraphics[width=0.92\textwidth]{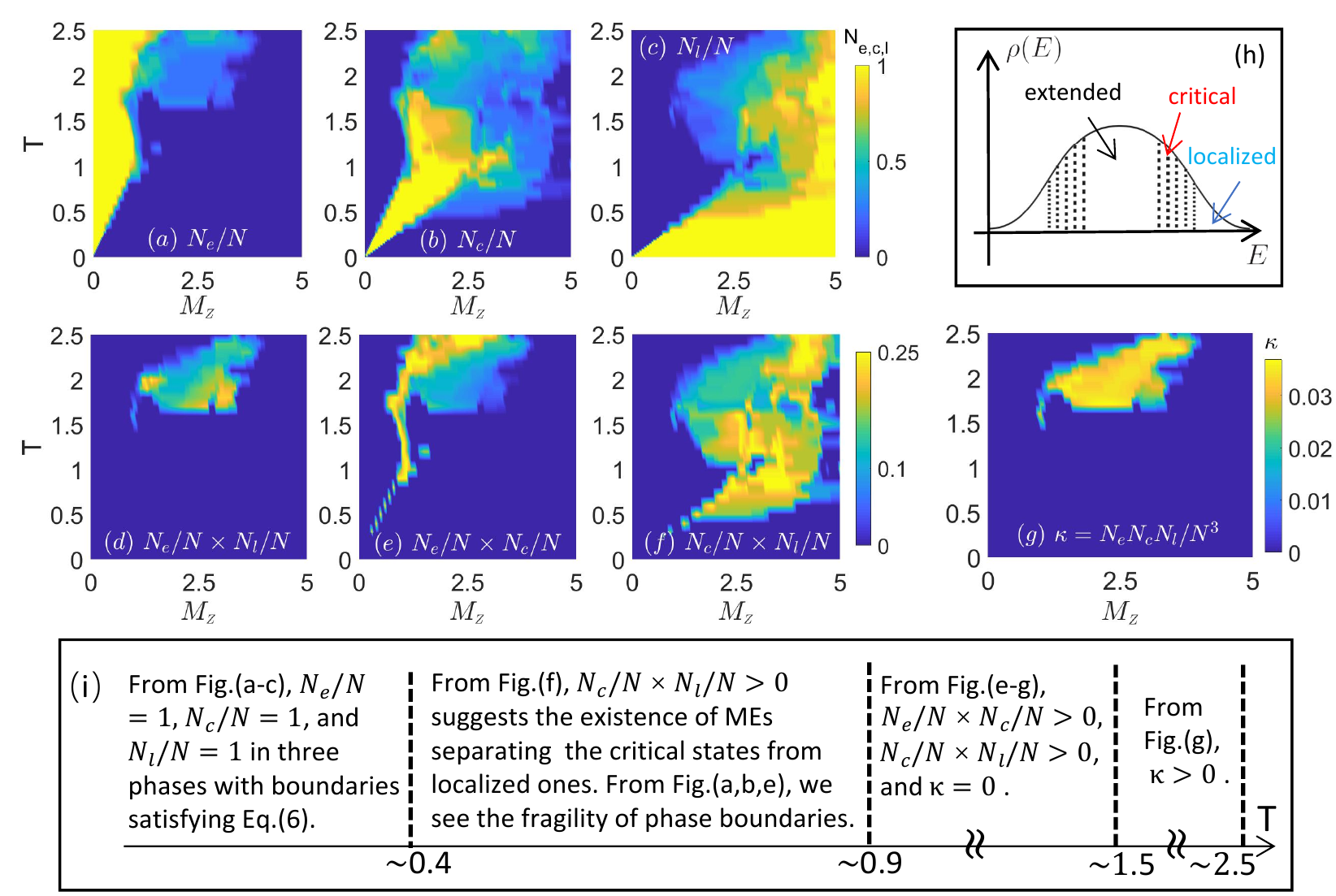}
\caption{\label{02}
 (a) $N_e/N$, (b) $N_c/N$, (c) $N_l/N$, (d) $N_e/N\times N_l/N$, (e) $N_e/N\times N_c/N$, (f) $N_c/N\times N_l/N$ and (g) $\kappa$ as a function of $M_z$ and $T$. (h) Schematic figure of the density of states $\rho(E)$ as a function of energy $E$ for a system with coexisting extended, critical, and localized regions. We note that (h) is a sketch map, and the positions of the three different regions depend on specific systems. Here we fix $J_0=1$ and $J_{so}=0.5$. From figures (a-c), we see that $N_e/N=1$, $N_c/N=1$ and $N_l/N=1$ in three phases with the boundaries satisfying Eq.~(\ref{boundary}) when $T<0.4$. With increasing $T$ to $T\in(0.4, 0.9)$, from figure (f), $N_c/N\times N_l/N>0$ corresponds to the phase with coexisting critical and localized states, which suggests the existence of MEs separating critical states from localized ones. In this region, from figures (a,b,e), we see $N_e=1$ and $N_c=1$ when the parameters slightly away from the boundary described as Eq.~(\ref{boundary}), but $N_e\times N_c>0$ at the boundary. The change occurs only at the region that is very close to boundary, meaning that the states in this region are  more susceptible. When $T\in(0.9, 1.5)$, there are the regions corresponding to $N_e/N\times N_c/N>0$ [figure (e)] or $N_c/N\times N_l/N>0$ [figure (f)], but no regions are $\kappa>0$, meaning that there exists the phase with coexisting extended (or localized) and critical regions, but no the phase with coexisting three different regions exists. When $T\in(1.5, 2.5)$, $\kappa>0$ means the existence of the phase with coexisting extended, critical and localized regions. We summarize these results in the figure (i).}
\end{figure*}
%%%%%%%%%%%%%%%%%%%%%%%%%%%%%%%%%%%%%%%%

To clearly and completely characterize the phase diagram of this system, we introduce the extended-state fraction $N_e/N$, localized-state fraction $N_l/N$, critical-state fraction $N_c/N$, and their product~\cite{YuchengC3},
\begin{equation}\label{kappa}
\kappa=\frac{N_e}{N}\times\frac{N_l}{N}\times\frac{N_c}{N},
\end{equation}
where $N_e$, $N_l$ and $N_c$ are the numbers of the extended, localized and critical eigenstates, respectively. These diagnostic quantities can characterize all different phases. $N_e/N=1$, $N_c/N=1$ and $N_l/N=1$ correspond to the extended, critical and localized phases, respectively. In the large $L$ limit, $N_e/N\times N_l/N >0$ and $\kappa=0$ characterize the conventional ME separating localized states from extended ones. $N_e/N\times N_c/N >0$ ($N_c/N\times N_l/N >0$) and $\kappa=0$ describes the ME separating critical states from extended (localized) states. The phase with coexisting localized, extended, and critical regions corresponds to $\kappa>0$.

Figs.~\ref{02} (a), (b) and (c) show the $N_e/N$, $N_c/N$ and $N_l/N$, respectively.
We see that when $T\ll 1$, this system possesses three phases with solely extended, critical and localized eigenstates, which correspond to $N_e/N=1$, $N_c/N=1$ and $N_l/N=1$, respectively, and the phase boundaries satisfy Eq.~(\ref{boundary}). With increasing $T$, eigenstates with different properties overlap each other. Figs.~\ref{02} (d), (e), (f) and (g) display the behavior of $N_e/N\times N_l/N$, $N_e/N\times N_c/N$, $N_c/N\times N_l/N$ and $\kappa$, respectively. We see that when $T>0.4$, there is a broad phase region corresponding to $N_c/N\times N_l/N>0$ and $\kappa=0$ [Figs.~\ref{02}(f) and (g)], meaning that there is the phase with coexisting critical and localized regions, and thus there are MEs separating the critical states from localized ones. Further increasing $T$, there will appear the phase with coexisting extended and critical regions [Fig.~\ref{02}(e)] and the phase with three coexisting regions [Fig.~\ref{02}(g)]. Fig.~\ref{02}(h) is a sketch of the phase with coexisting three fundamentally different regions, and one can see two types of MEs separating the localized and extended
regions from critical regions, respectively. Although there exists a broad region corresponding to $N_e/N\times N_l/N>0$ [Fig.~\ref{02}(d)], $\kappa$ is non-zero in this region [Fig.~\ref{02}(g)], meaning that there is not a phase with coexisting extended and localized regions but no critical region here. The different behaviors as the change of the driven period are summarized in Fig.~\ref{02}(i).
Further, in the range of $T\in (0.4, 0.9)$, from Figs.~\ref{02}(a-c,d,e), we see that the phases on both sides of the phase boundary between the extended and critical phases remain extended and critical, respectively, but the phases on either side of the phase boundary between the critical and localized phases are more easily influenced by the period $T$ and they no longer remain solely critical or localized. This phenomenon can be understood from Eq.~(\ref{Lambda6}), since $M_z$ near the phase boundary between the critical and localized phases is larger, which induces that the high-order terms are larger and impact the original phases more easily in the process of increasing $T$.

\section{two interesting phenomena: fragility of phase boundaries and reentrant localization transition}
\begin{figure}[t]
%\hspace*{-0.6cm}
\includegraphics[width=0.49\textwidth]{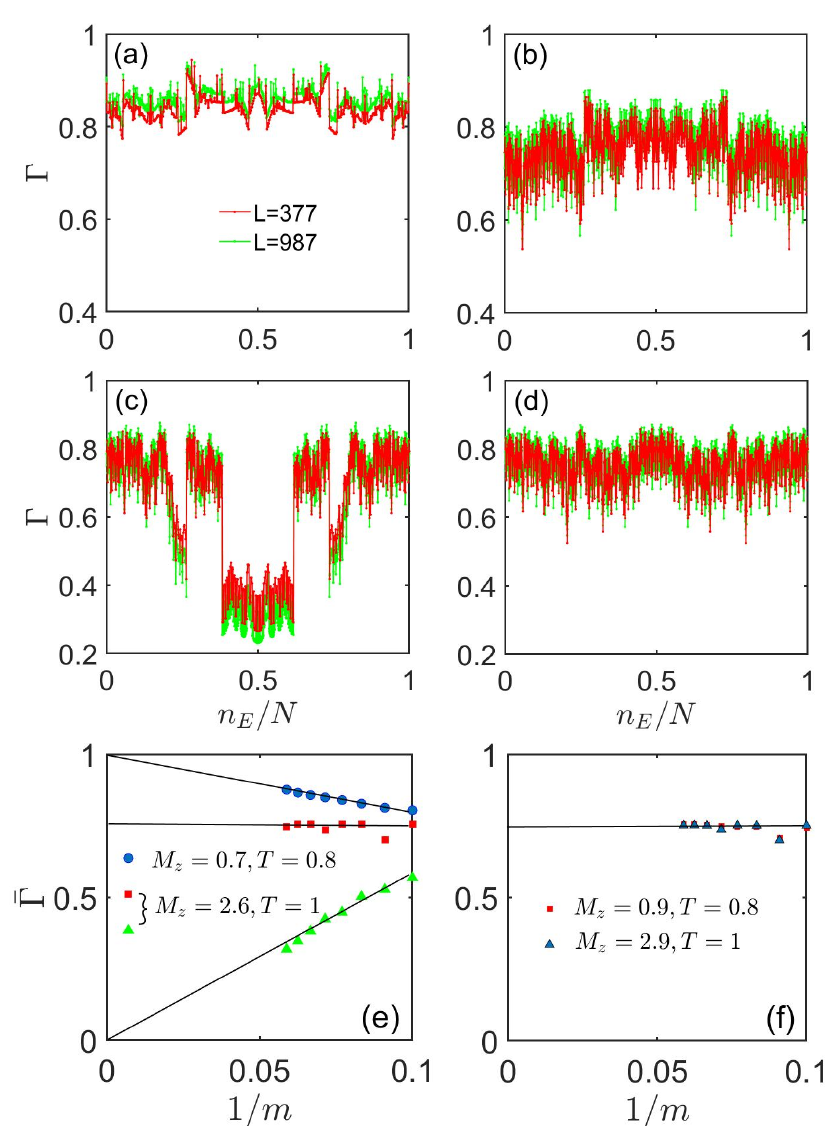}
\caption{\label{03}
 Fractal dimension $\Gamma$ as a function of $n_E/N$ for the sizes $L=377$ and $L=987$ with (a) $M_z=0.7$, $T=0.8$, (b) $M_z=0.9$, $T=0.8$, (c) $M_z=2.6$, $T=1$, and (d) $M_z=2.9$, $T=1$. (e) (f) $\bar{\Gamma}$ as a function of $1/m$. (e) The average $\Gamma$ over all eigenstates of (a) and different regions of (c). (f) The average $\Gamma$ over all eigenstates of (b) and (b). Here we set $J_0=1$ and $J_{so}=0.5$.}
\end{figure}

Besides the rich physical properties about the MEs and the critical phase or regions in the phase diagram, there are also two interesting phenomena.
From the above section, the phases on both sides of the extended-critical phases boundary are unaffected by the periodical kick when $T\in(0.4, 0.9)$, i.e., they remain the extended or critical behaviors. However, the states on the phase boundary are easily affected.
All eigenstates are originally critical on the phase boundary between the extended and critical phases when the quasiperiodic potential is non-kicked, i.e., $N_c/N=1$ and $N_e/N=0$ on the boundary. When $T\ll 1$, the effective Hamiltonian can be described by the Hamiltonian without the kicked case. Thus, the boundary is unaffected and has $N_c/N=1$. With increasing $T$, $N_e/N\times N_c/N$ becomes non-zero, as shown in Fig.~\ref{02}(e)~\cite{explainX}, which suggests that the boundary becomes from the situation with all eigenstates being critical to the situation with extended and critical states being coexisted. For the parameters slightly away from the boundary, the extended and critical phases remain unaffected. To illustrate this, we fix $T=0.8$ and show the fractal dimensions of all eigenstates with $M_z=0.7$ and $M_z=0.9$ in Figs.~\ref{03}(a) and (b), respectively. It can be seen that $N_e/N=1$ for $M_z=0.7$ [Fig.~\ref{03}(a)] and $N_c/N=1$ for $M_z=0.9$ [Fig.~\ref{03}(a)], which are slightly away from the boundary $M_{z}^{c}/T=2|J_0-J_{so}|=1$ (we have fixed $J_0=1$ and $J_{so}=0.5$) and show the similar properties with the non-kicked case. In comparison, on the boundary with $M_z=0.8$, $N_e/N\times N_c/N$ is larger than $0$, as shown in Fig.~\ref{01}(d), suggesting that the states are no longer solely critical.  Thus, the phase boundary is more susceptible to the periodical kick, which shows the fragility of the phase boundary.

%%%%%%%%%%%%%%%%%%%%%%%%%%%%%%%%%%%%%%%%%%%%%%%
%%%%%%%%%%%%%%%%%%%%%%%%%%%%%%%%%%%%%%%%%%%%%%%%

Another interesting phenomenon is the reentrant localization transition, namely that with increasing the quasiperiodic potential strength, after the AL transition, some of the localized
states become extended for a range of intermediate potential strengths, and eventually, these states undergo the second localization transition at a higher quasiperiodic potential strength~\cite{Basu2021}. Figs.~\ref{03}(c) and (d) show the fractal dimension of this system with $M_z=2.6$ and $M_z=2.9$ for the fixed $T=1$. We see that when $M_z=2.6$, there exist the critical and localized regions, but when $M_z=2.9$, all eigenstates become critical, meaning that with increasing the quasiperiodic potential strength, some localized states become delocalized. Naturally, further increasing the potential strength, these states once again become localized. The phenomenon of the reentrant localization transition can only occur when $T>0.8$, namely in the low-frequency region. We note that there is not the reentrant localization transition when the quasiperiodic potential is non-kicked~\cite{YuchengC2}, the occurrence of this phenomenon originates from that the potential is added in the kicked way.

To further confirm the extended, critical or localized properties in different regions, we carry out the finite size analysis by calculating $\bar{\Gamma}$, as shown in Figs.~\ref{03}(e) and (f). The mean fractal dimension $\bar{\Gamma}$ averaged over all eigenstates in Fig.~\ref{03}(a) tends to 1 [blue spheres in Fig.~\ref{03}(e)], suggesting that all eigenstates are extended. Similarly, we can confirm that the system in Fig.~\ref{03}(c) includes the localized and critical regions [red squares and green triangles in Fig.~\ref{03}(e)], and all eigenstates in Figs.~\ref{03}(b) and (d) are critical [see Fig.~\ref{03}(f)].

\
\

\section{experimental realization and detection}
%%%%%%%%%%%%%%%%%%%%%%%%%%%%%%%%%%%%%%%
\subsection{Experimental realization}
\begin{figure}[t]
%\hspace*{-0.8cm}
\includegraphics[width=0.45\textwidth]{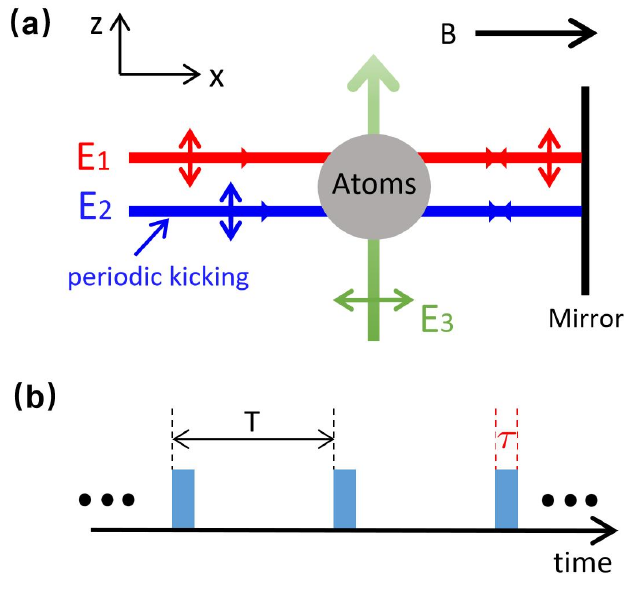}
\caption{\label{04}
(a) Schematic of the experimental setup. ${\bf E}_1$ is a standing wave with $z$ polarization, which generates the spin-independent main lattice.
${\bf E}_2$ is the kicked standing wave giving the kicked quasiperiodic potential. ${\bf E}_{3}$ is a plane wave, which is used to form the Raman coupling potential. (b) Experimental sequence composed of the finite-width, unit-height pulse with the effective pulse width $\tau$ and the pulse interval $T$.}
\end{figure}

We propose to realize the Hamiltonian (\ref{Hsum}) based on apodized Floquet engineering techniques~\cite{Shimasaki2022,Weld2022} and optical Raman lattices~\cite{LiuXJ2013,LiuXJ2014,Lepori2016,WangBZ2018,Liu2016,Song2018,Song2019,JWPan,RamanReview2018}. Fig.~\ref{04}(a) shows the schematic diagram, where ${\bf E}_1$ with $z$ polarization is a standing-wave beam and ${\bf E}_{3}$ with $x$ polarization
is a plane wave. They are applied to generate the spin-independent main lattice $V_1(x)=V_{\rm m}\cos^2({k}_1x)$ with the depth $V_{\rm m}$, which
induces the spin-conserved hopping ($H_0$), and a Raman coupling potential to generate the spin-flip hopping ($H_{\rm SOC}$).
The periodically kicked quasiperiodic potential potential ($H_K$) is realized by periodically applying another standing wave ${\bf E}_{2}$, which is used to generate a spin-dependent lattice $V_2(x)\sigma_x=V_{\rm s}\cos^2({k}_2x)\sigma_x$ with the depth $V_{\rm s}$.
In this setting, the lattice wave numbers $k_{1,2}$ are easily tunable in experiment and making them incommensurate to product the irrational number $\alpha=k_2/k_1$. In the tight-binding approximation, the realized Hamiltonian is given by

\begin{equation}\label{Hsumexp}
H=H_0+H_{\rm SOC}+F(t)\Delta\sum_{i}\cos(2\pi\alpha i)(n_{i,\uparrow}-n_{i,\downarrow}),
\end{equation}
with
\begin{equation}\label{HFt}
F(t)=\sum_ng_{\tau}(t-nT).
\end{equation}
being the waveform of the periodic pulse train. Here $g_{\tau}$ describes the shape of the pulse, $T$ is the pulse interval, and $\tau$ is the effective width of the single pulse: $\tau=\int^{\infty}_{-\infty}g_{\tau}(t)dt$, as shown in Fig.~\ref{04}(b), where we take the square pulses as an example. In the limit of small $\tau$~\cite{Shimasaki2022}, $\Delta F(t)=\sum_nM_z\delta(t-nT)$, where $M_z=\Delta\cdot\tau$, and then, the Hamiltonian (\ref{Hsumexp}) becomes the Hamiltonian (\ref{Hsum}).

%%%%%%%%%%%%%%%%%%%%%%%%%%%%%%%%%%%%%%%%%%%%%%%
%%%%%%%%%%%%%%%%%%%%%%%%%%%%%%%%%%%%%%%%%%%%%%%%

%%%%%%%%%%%%%%%%%%%%%%%%%%%%%%%%%%%%%%%%%%%%%%%
\begin{figure}[t]
\hspace*{-0.2cm}
\includegraphics[width=0.48\textwidth]{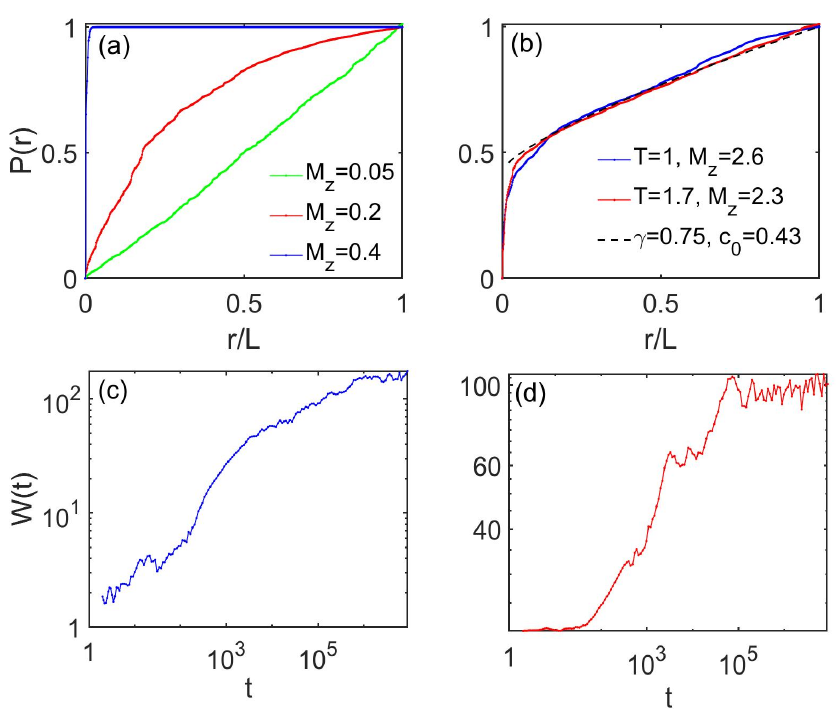}
\caption{\label{06}
Long-time survival probability ($t=5\times 10^7T$) with (a) $M_z=0.05$ (extended phase), $M_z=0.2$ (critical phase), $M_z=0.4$ (localized phase), and the fixed $T=0.1$, (b) $T=1$ and $M_z=2.6$ corresponding to the phase with coexisting critical and localized regions [see Fig.~\ref{03}(c)], and $T=1.7$ and $M_z=2.3$ corresponding to the phase with coexisting extended, critical and localized regions [see Fig.~\ref{01}(f)]. For (a) and (b), we take $20$ samples with a sample being specified by choosing an initial phase $\phi$. The dashed line is plotted by fixing $D_2=0.75$ and changing $c_0$ to fit the data points as Eq.~(\ref{Pr}). Log-log plot of $W$ versus the time $t$ for (c) $T=1$ and $M_z=2.6$, and (d) $T=1.7$ and $M_z=2.3$. We choose $L=987$ for all figures.}
\end{figure}
%%%%%%%%%%%%%%%%%%%%%%%%%%%%%%%%%%%%%%%%%%%%%%%%

\subsection{Experimental detection}
Next we study the detections of the different phases based on the expansion dynamics. We consider a wave packet with spin up initially at the center of the lattice, i.e., $|\psi(0)\rangle=c^{\dagger}_{(L+1)/2,\uparrow}|0\rangle$ (let the size $L$ be odd), and the final state is set as $|\psi(t)\rangle=\sum_{j=1}^L[u_{j}(t)c^{\dagger}_{j,\uparrow}+v_{j}(t)c^{\dagger}_{j,\downarrow}]|0\rangle$. We firstly focus on the survival probability $P(r)$ defined as
 \begin{equation}\label{SP}
P(r)=\sum_{|j-\frac{L+1}{2}|\leq r/2}|(u_{j}(t)|^2+|v_{j}(t)|^2),
\end{equation}
which describes the probability of finding the particle after a given time $t$ in the sites within the region $[-r/2, r/2]$~\cite{Santos2019}. After a long time evolution ($t\rightarrow\infty$), $P(r)$ is proportional to $(r/L)^{D_2}$ with $D_2$ being the generalized dimension of spectral measures~\cite{Santos2019,Geisel1992,Xu2020}. For the extended phase, the distribution of the final state will be uniform, and thus, $P(r)$ linearly increases as $r$ increases. For the localized phase, the particle will localize at the position near the initial point, and thus, $P(r)$ quickly reaches $1$ within a small $r$. For the critical phase, the distribution is delocalized and nonergodic, and thus, $P(r)$ reaches $1$ when $r\rightarrow L$ but the increasing rate is not linear. Fig.~\ref{06}(a) shows the typical distributions of $P(r)$ with long times in the extended phase (green line), critical phase (red line) and localized phase (blue line). For sufficiently large $r/L$, we have $P(r)\approx (r/L)^{D_2}$ with $D_2=0, 1$ and $0<D_2<1$ for the localized, extended and critical phases, respectively, and $0<D_2<1$ reflects the nonergodic character of the critical phase. For a system with MEs, the distribution of $P(r)$ will become complex.
Fig.~\ref{06}(b) shows the $P(r)$ of the phase with coexisting localized and critical regions (blue line) and the phase with coexisting three different regions (red line). $P(r)$ dramatically increases for a small $r$, suggesting the existence of localized regions, but reaches $1$ when $r\rightarrow L$, meaning that there also exist the delocalized regions. For the phase with coexisting localized and critical regions, the increasing rate of $P(r)$ is determined by the states in the critical region, and the average fractal dimension can be extracted by
\begin{equation}\label{Pr}
P(r)=(r/L)^{D_2}(1-c_0)+c_0,
\end{equation}
where $c_0$ is the constant that depends on the proportion of the localized states in all eigenstates. Fig.~\ref{03}(e) tells us $\bar{\Gamma}\approx0.75$, and thus we plug $D_2=0.75$ into Eq.~(\ref{Pr}) to well fit the $P(r)$. From Fig.~\ref{06}(b), Eq.~(\ref{Pr}) with $D_2=0.75$ can also be well fit to the coexisting phase with three different regions. Thus, it is difficult to further distinguish whether the delocalized regions are critical or the co-existing of critical and extend regions from $P(r)$ with $t\rightarrow\infty$.

To see the differences between the two cases in Fig.~\ref{06}(b) in dynamics, we should not consider the distributions after a long time evolution. Instead, we should consider the process of the expansion of the wave packet. To characterize the expansion of the above initial state, we consider the mean square displacement~\cite{YuchengC2,Hiramoto,Ketzmerick,Roati2008},
\begin{equation}\label{MSD}
W(t)=\sqrt{\sum_{j}[j-(L+1)/2]^2 (u_{j}(t)|^2+|v_{j}(t)|^2)},
\end{equation}
which measures the width of the wave packet after the evolution time $t$. $W(t)$ can be expressed as $W(t)\sim t^{\gamma}$ with $\gamma$ being the dynamical index. For AA model, $\gamma=0$, $\gamma=1$ and $\gamma\approx\frac{1}{2}$ in the localized phase,
extended phase and critical point, respectively, meaning that the corresponding expansion is localized, ballistic,
and normal diffusive, respectively. For the coexisting phase, $W(t)$ is not straightforward to $t^{\gamma}$, as shown in Fig.~\ref{06}(c) and (d). It is obvious that the coexisting phase including the extended region expands more quickly and reaches the boundary faster.
Further, from the viewpoint of the transport~\cite{Purkay,Saha2019}, the conductivity is independent of the system size in the extended region, while it decreases in the power-law and exponential fashion with the system size in the critical and localized region, respectively. Thus, by shifting the position of the Fermi energy across different regions, one can detect the corresponding transport properties, and further obtain the more precise information of the coexisting phases.

%%%%%%%%%%%%%%%%%%%%%%%%%%%%%%%%%%%%%%%%
\section{summary}
%\label{non-interacting}
%%%%%%%%%%%%%%%%%%%%%%%%%%%%%%%%%%%%%%%
We have investigated the critical and localized properties in the 1D periodically kicked quasiperiodic optical Raman lattice by comparing the fractal dimensions with different sizes. This system shows a rich phase diagram. In the high frequency regime ($T\ll 1$), the transition between the extended and critical phases occurs at $M_z/T=2|J_0-J_{so}|$, and the transition between the critical and localized phases occurs at $M_z/T=2(J_0+J_{so})$, which can be interpreted from the effective Hamiltonian of this system. With increasing $T$, there are the phase with coexisting critical and localized regions, and the phase with coexisting extended and critical regions. The two phases exhibit two types of MEs which separate the localized states from critical ones, and the extended states from critical ones, respectively. Further increasing $T$, there is the coexisting phase of extended, critical and localized regions. We have also found the fragility of the phase boundary, namely that the phase boundary is more susceptible to the dynamical kick, and the phenomenon of the reentrant localization transition. Finally, we have studied in detail the experimental realization, which can be immediately achieved in the current experiments, and the experimental detection based on the expansion dynamics of the wave packet.
Our results show that the periodically kicked incommensurate optical lattice is a new effective way to study and detect the novel critical phase, MEs, coexisting quantum phases and some other interesting phenomena.

\begin{acknowledgments}
We thank C. Yang and Y. Peng for reading our manuscript carefully. This work is supported by National Key R\&D Program of China under Grant No.2022YFA1405800, the National Natural Science Foundation of China (Grant No.12104205), the Key-Area Research and
Development Program of Guangdong Province (Grant No. 2018B030326001), Guangdong Provincial Key Laboratory (Grant No.2019B121203002).
\end{acknowledgments}
%This will simplify the ME's research in both theory and experiment. For example, in experiment, it is difficult to accurately realize the Model I (\ref{ME1}) and detect the location of ME, but owing to the dual relation, the above-mentioned work~\cite{Gadway2020} that realized the Model II (\ref{ME2}) in momentum space can be considered to have also realized the Model I in real space and detected the location of ME.
%\clearpage

%%%%%%%%%%%%%%%%%%%%%%
\appendix
%%%%%%%%%%%%%%%%%%%%%%%%%%%%
\section{Details for the finite size scaling analysis}
In the main text, we carried out the finite size scaling analysis by plotting the $\bar{\Gamma}$ as a function of $1/m$. Since we focused on the value of $\bar{\Gamma}$ when the system size tends to infinity and used it to distinguish different regions, we omitted some details, such as error bars or uncertainty analysis of the fit. Before discussing these details, we firstly explain why we used the fractal dimension instead of the inverse participation ratio (IPR), even though IPR is the frequently-used quantity to distinguish localized from extended states. For extended
states, one has $IPR\sim 1/L \rightarrow 0$ for $L\rightarrow\infty$, and for localized states the IPR approaches a finite nonzero value for $L\rightarrow\infty$. Therefore, the IPR-criterion can distinguish only between localized and extended states, there is no
room for a third kind of states. Therefore, we use the fractal dimension $\Gamma$, which tends to $0$, $1$ and a finite value between $0$ and $1$ when $L\rightarrow\infty$ for localized, extended and critical states, respectively.

%%%%%%%%%%%%%%%%%%%%%%%%%%%%%%%%%%%%%%%%%%%%%
\begin{figure}[t]
\hspace*{-0.3cm}
\includegraphics[width=0.49\textwidth]{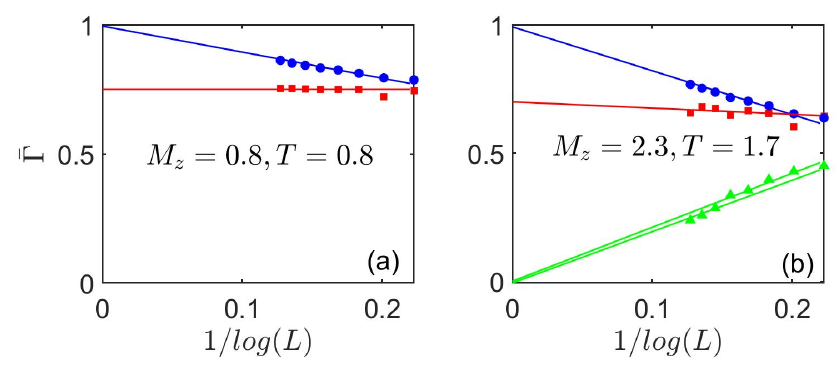}
\caption{\label{07}
$\bar{\Gamma}$ as a function of $1/log(L)$ for different regions with (a) $M_z=T=0.8$ (corresponding to Fig.~\ref{02S} (a)), and (b) $M_z=2.3$, $T=1.7$ (corresponding to Fig.~\ref{02S} (b)). (a) The blue line is $y=-1.01x+0.995$ and the red line satisfies $y=0.75$. (b) The blue line is $y=-1.71x+0.992$, the red line is $y=-0.243x+0.699$, and the two green lines respectively satisfy $y=2.0x-0.006$ and $y=2.1x+0.001$.}
\end{figure}
%%%%%%%%%%%%%%%%%%%%%%%%%%%%%%%%%%%%%%%%%%%%%%%%

There are variety of ways to perform the finite size scaling analysis. For example, the horizontal axis can be $1/log(L)$ or $1/m$, where $m$ is the Fibonacci index, and the ordinate axis can be the fractal dimension or index $\alpha_{min}$, where $\alpha_{min}=-\frac{log(n_{max})}{log(L)}$ with $n_{max}$ being the distribution peak~\cite{YuWang2020,WangJ2016,TLiu2022,YuchengC2}. Fig.~\ref{07} show $\bar{\Gamma}$ as a function of $1/log(L)$ for different regions with the same parameters in Fig.~\ref{02S}. We use the linear function $y=Ax+B$ to fit the data points, where $A$ and $B$ are the undetermined coefficients. One can determine $A=-1.0102\pm 0.1208, B=0.9952\pm 0.0482$ and $A=-0.13825\pm 0.22405, B=0.7751\pm 0.0394$ for the blue and red data points in Fig.~\ref{07} (a), and $A=-1.71635\pm 0.17335, B=0.992\pm 0.0267$ and $A=-0.45305\pm 0.55795, B=0.7288\pm 0.0951$ for the blue and red data points in Fig.~\ref{07} (b), and $A=2.1180\pm 0.2155, B=0.0011\pm 0.0178$ and $A=2.09175\pm 0.20375, B=-0.00655\pm 0.03385$ for the green data points in Fig.~\ref{07} (b). One can see that there exist obvious deviations for the second red data point from the right in both (a) and (b), which lead to larger errors of the fitting results. After removing the second red data point in the process of fitting, we can determine $A=-0.05498\pm 0.0298, B=0.7539\pm 0.0050$ for the red data points in Fig.~\ref{07} (a), and $A=-0.243\pm 0.3865, B=0.6996\pm 0.064$ for red data points in Fig.~\ref{07} (b). We see that when $L\rightarrow\infty$, the $\bar{\Gamma}$ of the extended and localized regions respectively tend to $1$ and $0$ within error permissibility, while for the critical region, $\bar{\Gamma}$ is far from $0$ and $1$, manifesting that the critical states are fundamentally different from the extended and localized states. The same kind of analysis applies to the case that the horizontal axis is $1/m$.

%%%%%%%%%%%%%%%%%%%%%%%%%%%%%%%%%%%%%%%%%%%%%%


\begin{thebibliography}{54}
\bibitem{explainX} The discontinuities of $N_e/N\times N_c/N>0$ on the phase boundary in Fig.~\ref{02}(e) originate from the existence of the data interval in the calculation process, and thus some parameters on the boundary are not covered.
\end{thebibliography}
\end{document}